# Nonlinear ultrafast fiber amplifiers beyond the gain-narrowing limit

PAVEL SIDORENKO,* WALTER FU, AND FRANK WISE

*School of Applied and Engineering Physics, Cornell University, Ithaca, New York 14853, USA,*
*\*Corresponding author, E-mail: ps865@cornell.edu*

Ultrafast lasers are becoming increasingly widespread in science and industry alike. Fiber-based ultrafast laser sources are especially attractive because of their compactness, alignment-free setups, and potentially low costs. However, confining short pulses within a fiber core leads to high intensities, which drive a host of nonlinear effects. While these phenomena and their interactions greatly complicate the design of such systems, they can also provide opportunities for engineering new capabilities. Here, we report a new fiber amplification regime distinguished by the use of a dynamically-evolving gain spectrum as a degree of freedom: as a pulse experiences nonlinear spectral broadening, absorption and amplification actively reshape both the pulse and the gain spectrum itself. The dynamic co-evolution of the field and excited-state populations supports pulses that can broaden spectrally by almost two orders of magnitude and well beyond the gain bandwidth, while remaining cleanly-compressible to their sub-50-fs transform limit. Theory and experiments provide evidence that a nonlinear attractor underlies the management of the nonlinearity by the gain. Further research into the mutual, pulse-inversion propagation dynamics may address open scientific questions and pave the way toward simple, compact fiber sources that produce high-energy, sub-30-fs pulses.

## Introduction

High-peak-intensity ultrafast fiber lasers are desirable for industrial, defense, and scientific applications[1–3]. Fiber lasers possess many advantages over their solid-state counterparts including compactness, good thermo-optical properties, and excellent beam quality. On the other hand, the confinement of high-peak-power pulses in a small core for long propagation distances results in the accumulation of large nonlinear effects, which degrade the pulse quality. High-energy sources are often based on chirped-pulse amplification (CPA)[4], in which the pulse from an oscillator is stretched, amplified, and subsequently compressed. Stretching reduces the peak intensity, and in turn, the strength of nonlinear effects such as self-phase modulation (SPM) and stimulated Raman scattering (SRS). In a typical fiber CPA system, the pulse duration is limited to ~200 fs by gain narrowing[1] and residual dispersion mismatch between the stretcher and compressor. Although the dispersion mismatch can be mitigated to some degree by SPM-induced nonlinear phase[5], gain narrowing remains an ongoing and major challenge. While 200-fs pulses are suitable for some applications, others, such as high harmonic generation, benefit from shorter pulse durations[6]. Shorter pulses can be obtained by a subsequent stage of nonlinear pulse compression, which increases the complexity of the system and decreases the overall efficiency[7]. In contrast to CPA, there exist techniques based on nonlinear pulse propagation that can be used to achieve sub-100-fs pulse durations. These include direct amplification[8] and pre-chirp management[9,10] as well as amplifiers designed to support self-similar pulse evolution[11–13]. While such approaches can yield sub-100-fs pulses[9,14,15], they suffer from their own set of limitations. Similariton amplifiers benefit from a nonlinear attractor[16] that is indifferent to many of the seed characteristics, but at higher energies their spectra overflow the gain bandwidth, degrading the compressed pulse quality[17]. Nonlinear amplifiers using pre-chirp management can extend this limit and reach energies as high as the microjoule level and durations as short as 24 fs[15]; however, achieving optimal performance often requires a carefully-chosen seed pulse, which hampers energy scaling and increases the complexity of the design.

Here, we describe and experimentally demonstrate a new regime of fiber amplification that addresses two main challenges of ultrafast fiber lasers: 1) management of the high nonlinear phase shifts encountered in stretcher-less amplification systems , and 2) generation of bandwidths much broader than the gain spectrum that can be compressed to clean, sub-100-fs pulses. In this regime, the pulse and gain spectra evolve and reshape one other in tandem, so we refer to this as amplification with gain-managed nonlinearity (GMN). The intentional use of co-evolving gain to control pulse evolution is rare. In contrast to previously-investigated amplifiers where doping was used as a degree of freedom to reduce the nonlinearity and heat load[18,19], we engineer the evolution of the gain spectrum to deliberately take advantage of nonlinear effects. In this regime, an accumulation of nonlinear phase on the order of ~200π (which corresponds to 70-fold spectral broadening) can be accommodated while still generating clean compressed pulses. Experimentally, we demonstrate the GMN regime in ytterbium-doped fiber and obtain broadband amplified pulses that can be compressed to ~40 fs durations. Numerical simulations and experimental results suggest that the pulse evolution in the gain medium is driven by a nonlinear attractor, which has interesting scientific implications and may simplify device design.

## Results
### Numerical simulations

We begin by illustrating the GMN amplification regime numerically by a simple example. Figure 1 depicts the simulated pulse evolution in a typical, highly-doped, ytterbium fiber amplifier. A 0.5-nJ, 0.8-ps Gaussian pulse at 1028 nm is launched into an Yb-doped fiber amplifier with a 6-μm core diameter, which is co-pumped at 976 nm. The numerical model includes second- and third-order dispersive effects, along with self-phase modulation, self-steepening, Raman scattering, and the gain calculated by solving the population inversion rate equations simultaneously with the pulse evolution[20,21]. The pulse evolution initially follows a well-known trajectory: dispersive and nonlinear effects are both prevalent, and the pulse quickly broadens into a self-similar pulse or similariton[12,13] (Fig. 1b). As the pulse's energy and bandwidth increase, gain-shaping begins to impede the similariton's growth, leading to the expected nonlinear pulse distortions and the loss of self-similar propagation[17,22] (Fig. 1c). This point is traditionally taken to be the limit of self-similar amplification, with the result that such systems rarely reach useful energies for high-performance applications. However, if the pulse evolution is allowed to continue, a new regime emerges. As the inversion decays away, the pulse shifts towards longer wavelengths and continues broadening in both the spectral and temporal domains. The pulse energy increases and the pulse evolves to an asymmetric temporal profile with a smooth spectrum (Fig. 1d). Perhaps most surprisingly, the new pulse develops a monotonic chirp that can be compensated with a grating pair (as will be shown below).

This new regime that unexpectedly emerges beyond self-similar amplification is of obvious interest from the standpoints of both nonlinear wave physics and applications. To understand its origins, we perform additional, more detailed, simulations. We find that modeling the true gain calculated from the rate equations is of crucial importance, as the quasi-three-level nature of Yb-doped fiber leads to the gain spectrum changing dynamically as different components of the pump and signal waves are absorbed and reemitted. This leads to a longitudinally-evolving gain spectrum and an intertwining of the roles of nonlinearity and gain-shaping, with the end result that the nonlinear phase is managed and a compressible frequency chirp is maintained. The main features of the GMN amplification regime are illustrated in Figures 2 and 3. We find that the mutual interaction of the signal and gain can be enhanced in long, highly-doped gain fibers, and when the seed wavelength is chosen to be close to the transition between absorption and gain in the fiber. We therefore keep the same 1028-nm seed wavelength as before, but increase the transform-limited duration to 2 ps in order to explore the use of different seed pulses. As above, a Yb-doped fiber amplifier with a 6-μm core diameter that is co-pumped at 976 nm is assumed. Initially, the pulse evolution is dominated by SPM (Figs. 2a-b). Remarkably, after ~2 m, the pulse begins evolving toward a smooth, asymmetric temporal shape instead of experiencing the expected wave-breaking distortions[23,24]. The longitudinal evolution of the gain spectrum is shown in Figures 2a-e by the solid red curve. The GMN regime is achieved when nonlinear spectral broadening balances gain-shaping. This balance manifests as absorption of the blue part of the spectrum and amplification of the red part (Figs. 2c-e, 1j-2, and 3a). This deliberate use of both gain and loss during pulse amplification is a crucial aspect of the GMN regime, despite running counter to conventional amplifier design rules. Furthermore, unlike every other established amplification regime, the peak power of the pulse does not increase monotonically with propagation through the amplifier. Instead, it reaches a maximum at ~2 m and decreases thereafter (Figs. 2h-l and 3b). This nonmonotonic behavior is due to the dynamic balance between gain and dispersion, as can be seen in Figures 3b and 3c. During the initial stage of evolution, where the pulse is evolving toward the parabolic attractor, the peak power grows as expected due to the nearly-exponential, broadband (relative to the pulse) gain, and the pulse bandwidth increases. The effects of dispersion increase due to the nonlinearly-accruing bandwidth, while the rate of amplification decreases as more of the pump is absorbed; after ~2 m of propagation, dispersive temporal broadening overcomes the gain and causes the peak power to subsequently decrease. The pulse energy continues to increase monotonically, aided by the increasing temporal duration, while the pulse maintains its time-domain intensity profile.

Close comparison of the spectral evolution of the pulse and gain helps to understand the GMN regime. The initial spectral broadening is arrested when the blue side of the spectrum encounters the absorption near 1020 nm (Figs. 3a-c). This absorption feature and the gain window both shift to the red with continuing evolution, so the blue side of the pulse spectrum is limited while its red side continues to broaden unhindered. As the peak power and the net gain start to decay, the growth of the pulse spectral bandwidth saturates, hemmed in by residual absorption on the blue side and vanishing gain on the red side. This is the origin of the asymmetric, steady-state pulse that emerges at the output of the amplifier (Figs. 2f and 4a). Surprisingly, following this complicated and dynamic mixing of spectral components, the pulse (Fig. 4a) can be easily-compressed to near the transform limit (Fig. 4b). The deviation from a linear chirp is readily-accommodated by the standard design of a grating compressor[25]. Thus, despite the extremely nonlinear and broadband pulse evolution, a simple grating compressor suffices to produce a high-quality, nearly transform-limited pulse. The cleanliness of this compression despite $120\pi$ of nonlinear phase accumulation is a hallmark of similariton evolution; however, the pulse has clearly developed beyond the similariton regime in terms of its propagation trajectory, its bandwidth, and its overall performance. According to our simulations this pulse evolution requires only the presence of group-velocity dispersion, self-phase modulation, and the true gain as calculated from the rate equations.

The simulations also provide indications that the final, asymmetric pulse shape is a nonlinear attractor. That is, the pulse evolution in the GMN regime is driven more by nonlinearity and gain-shaping than by the particular parameters of the seed pulse. To demonstrate this, we simulate launching seed pulses that span a range of shapes, energies, durations, chirps, bandwidths, and peak powers (Fig. 5a) into a fixed amplifier. The different pulses evolve in a qualitatively-similar manner: each passes through an initial, SPM-dominated transient before entering the GMN amplification regime. The transition between these two regimes depends on the seed pulse, and can be approximated based on the peak power, which reaches a maximum near the transition point and subsequently decreases as is characteristic of the GMN regime (Fig. 5b). In particular, although the different pulses evolve at different rates, we can compare them at equivalent points in their evolutions by noting where each crosses a peak power threshold (dashed black line in Fig. 5b). At that point, the pulses are nearly identical (Fig. 5c), providing strong evidence that the amplified pulse is a nonlinear attractor.

Scientifically, nonlinear attractors are interesting objects to study due to their fundamental role in nonlinear dynamical systems. To our knowledge, the only other nonlinear attractor in normal-dispersion amplifiers is the similariton[12,13]. The rarity of nonlinear attractors in fiber systems and the similarities between the two regimes make it natural to wonder whether the similariton and GMN regimes might be related. We currently believe that they are not, due to several features which fundamentally distinguish the GMN regime. It is well-known that, although similaritons can exist within various homogeneously-broadened[12] and saturated[26] gain models, they will distort and become incompressible in the presence of gain-shaping[17,22]. In contrast, the GMN regime can *only* exist when strong gain-shaping is present, with the effects of both gain and loss included. This is exemplified by the spectrum in Figure 2, which significantly exceeds the gain bandwidth (red curves in Figs. 2a-e) over much of the pulse's propagation. Furthermore, a similariton is characterized by exponential growth in its energy, duration, peak power, and bandwidth. A pulse in the GMN regime grows only in duration (Fig. 3c), while its energy and bandwidth saturate (Figs. 3b-c) and its peak power decreases (Fig. 3b). However, further research into the nature of the GMN regime may reveal connections underlying these two attractors.

**Experimental results**

Guided by numerical simulations, we experimentally demonstrate GMN amplification in 5 meters of highly-doped Yb fiber with 5-μm core diameter (Nufern PM-YDF-5/130-VIII), co-pumped by a 976-nm diode. We seed the amplifier with transform-limited, 700-fs, 1-nJ pulses. With increasing pump power, the spectrum broadens substantially, in agreement with simulations (Fig. 6a). The measured bandwidth rapidly grows to extend past 1100 nm, overflowing the bandwidth of Yb-doped gain media. Using a grating compressor, we are able to dechirp the amplified pulses to their transform limit, indicating that the generated bandwidth is coherent rather than originating from shot-noise-seeded Raman scattering The coherence of the pulses across their full bandwidths is further confirmed using dispersive Fourier transform measurements[27] (not shown), which show that the single-shot pulse spectra are stable and low-noise. For pulse energies of 68 nJ or higher, Raman scattering is observed, but only as a low-intensity spectral shoulder that is clearly distinguishable from the main pulse. (Quantitative discrepancies between the simulated and experimental Stokes waves are discussed in supplementary information S1.) At the highest output pulse energy (107 nJ, where the Raman contribution is less than 3% of the pulse energy), the amplified pulses are compressed to 42 fs, near the transform limit.

To investigate the numerical prediction that the pulse evolves to a nonlinear attractor, we launch a variety of seed pulses into an amplifier. In one experiment, the seed spectrum and duration are held constant while the energy is varied by nearly a factor of 3 (Figs. 7a-b); in another, a filter is used to vary the bandwidth of the transform-limited seed (and thereby, vary its duration) while its energy is held constant at 0.5 nJ (Figs. 7c-d). With either type of variation, the output spectrum (Figs. 7a,c) and compressed pulse (Figs. 7b,d) remain largely invariant, with no adjustment of the pump power or grating compressor. This stark insensitivity of the output to the seed characteristics is a strong indication that a nonlinear attractor underlies the GMN amplification regime.

Considering the significance of the attracting pulse shape in the GMN regime, it is important to measure the pulse directly from the amplifier and compare it with simulations. Unfortunately, our frequency-resolved optical gating (FROG)[28] instrument lacks the scanning range to measure the chirped pulse from the amplifier described above. Therefore, we constructed a shorter amplifier, seeded by narrowband transform-limited pulses with an energy of 3 nJ. This yields shorter, lower-energy, less-broadband amplified pulses that can be measured directly with FROG. Simulations of this amplifier exhibit good agreement with the measurements (Fig. 8). We limited the pulse energy to 77 nJ to limit the bandwidth. The measured output pulse exhibits the expected asymmetric temporal shape and good agreement with simulations (Fig. 8b), and can be compressed to near the transform limit (Fig. 8c).

**Discussion**
In this work, we have presented initial demonstrations of the GMN amplification regime. Several aspects of this work present opportunities for future investigations.

Although our focus has been on the nonlinear-wave physics, it should be mentioned that the results obtained here already offer performance advantages over prior approaches. The generation of stable 200-nJ and 40-fs pulses (Fig. S2 in Supplementary Information) with ordinary, single-mode fiber is remarkable. Simulations indicate that 300-nJ and 30-fs pulses should be possible with greater pump power. It will also be interesting to investigate pulse energy scaling in the GMN regime to large mode area (LMA) fibers: for double-clad LMA fibers, the rate of pump absorption will affect the longitudinal gain evolution in a non-trivial manner, which may result in scaling relations more complex than simple mode-area scaling. While we have focused on the short pulses obtainable using this approach, filtering the output yields energetic, transform-limited pulses in spectral regions outside conventional gain windows (Supplementary Information S5).

The GMN regime exists in systems where a longitudinally-varying population inversion gives rise to a changing gain spectrum. While we have focused on Yb-doped fiber as an exemplary case of this regime, other widely-used gain media such as Er-, Tm-, and Nd-doped fibers also satisfy this criterion. Thus, the GMN regime should be accessible in other spectral regions, provided that normal dispersion can be obtained. An example of an Er-based GMN amplifier is shown in Fig. S3. It will also be interesting to look into systems where the gain and loss are artificially arranged. For example, we might envision fiber Bragg gratings written directly into gain fibers[29] being used to manipulate and engineer the evolution of the gain and loss[30,31]. Combinations of different gain media, such as Yb-Er co-doped gain fibers, and/or the simultaneously use of co- and counter-propagating pump fields, could also be employed to further engineer the longitudinal gain evolution.

The possible existence and nature of a nonlinear attractor motivate future work. So far, we have been unable to derive analytical solutions of the nonlinear Schrödinger equation coupled with the population rate equations. We expect that an analytical solution will reveal more properties and provide a better understanding of the GMN amplification regime. As an example of a practical application of the nonlinear attractor, it may be possible to reduce or compensate noise or drift in a source of seed pulses. This would allow even relatively-noisy sources, such as harmonically mode-locked oscillators[32], to be used to seed GMN amplifiers. Investigation of the noise properties of GMN amplifiers will be a subject of future work that may open a way for noise reduction in fiber amplifiers. Numerical simulations (Supplementary Information S4) show that the GMN regime underlies the pulse evolution in recently-developed high-performance fiber oscillators[33,34]. We anticipate that a better understanding of the nature of the nonlinear attractor in the GMN regime will significantly advance the design and optimization of such oscillators.

From the standpoint of performance, it is interesting to consider how GMN-based fiber amplifiers might be scaled to extremely large mode areas. Such fibers exist[35], but it is well-known that the existence and population of multiple transverse modes can lead to reductions in the output beam quality[36]. Kerr beam self-cleaning[37] is a recently-discovered technique that shows promise as a means of improving the quality of multimode beams, but relies on strong Kerr nonlinearity. The ability of the GMN regime to tolerate similarly high nonlinear phase shifts raises the intriguing possibility of combining these two techniques: one might envision GMN-based fiber amplifiers that can take advantage of huge-core, highly-multimode fibers without sacrificing beam quality. If realized, such systems could extend fiber lasers to the 100-MW level while retaining ~30-fs pulse durations, putting them for the first time in direct competition with solid-state CPA systems.

Most broadly, nonlinear Schrödinger equations play an important role in science by explaining a variety of nonlinear phenomena in physics[38,39]. Thus, results presented here may influence advances not only in the field of nonlinear optics but also in areas such as superconductivity, plasma physics, and nonequilibrium statistical mechanics, where many phenomena are governed by nonlinear Schrödinger equations.

## Conclusion

In conclusion, we numerically and experimentally demonstrate a new amplification regime in which nonlinear spectral broadening is balanced by strong, longitudinally-evolving gain-shaping. In contrast with conventional ultrafast amplifiers, in which the gain is static, in the GMN regime, the evolution of the gain spectrum is a complementary and previously-unexplored degree of freedom. The distinct characteristic of this regime is a nonlinear attractor featuring extreme spectral broadening well beyond the gain bandwidth while producing pulses that can be compressed to nearly the transform limit.

## Materials and methods
### Numerical simulations

In this work numerical simulations where performed by modeling the pulse evolution with the generalized nonlinear Schrödinger equation (GNLSE). When accounting for the effects of gain, dispersion, SPM, self-steepening and stimulated Raman scattering, the GNLSE can be written in the following form[40]:

$$\frac{\partial A}{\partial z} = -\frac{\alpha(z,\omega)}{2}A - \left(\sum_{n\geq 2}\beta_n \frac{i^{n-1}}{n!}\frac{\partial^n}{\partial T^n}\right)A + i\gamma(1 + \frac{1}{\omega_0}\frac{\partial}{\partial T})((1-f_r)A|A|^2 + f_r A\int_0^\infty h_r(\tau)|A(z,T-\tau)|^2 d\tau)$$

where $\alpha$ is the spectrally dependent gain/loss coefficient and $\beta_n$ are the higher-order dispersion coefficients obtained by a Taylor series expansion of the propagation constant $\beta(\omega)$ around the center frequency $\omega_0$. The fractional contribution of the delayed Raman response to the nonlinear polarization is represented by $f_r$. In this work $f_r = 0.245$ and the analytical form of the delayed Raman response $h_r(t)$ was taken from Ref [41]. For numerical integration of the GNLSE we used the fourth-order Runge–Kutta method[42] in the interaction picture[43]. For gain/loss calculations we used a combined model that couples the laser rate equation and the GNLSE[20,44]. In particular, an iterative procedure of gain/loss calculations derived in Ref. [21] was used in all numerical results presented in this paper.

### Experiments
The seed source for the experiments was a normal-dispersion fiber oscillator[45] operating at 24 MHz repetition rate. The central wavelength of the oscillator was around 1030 nm and the pulse energy was 2.5 nJ. The pulse from the oscillator was compressed and spectrally filtered before the amplification. For the results presented in Figs. 6-7, the amplifier was constructed from 5 meters of Yb-doped fiber (Nufern PM-YDF-5/130-VIII). For the results presented in Figs. 7c-d, a spectral filter before the amplifier was used to generate different seed pulses. For the results presented in Fig. 8, the amplifier was constructed from 3 meters of Yb-doped fiber (Nufern PLMA-YDF-10/125-VIII). For pulse compression, we used a grating compressor with transmission gratings (Lightsmyth T-1000-1040). Temporal pulse characterization was done with SHG-FROG (FROGscan from Mesaphotonics), and a ptychographic algorithm[46] was used to reconstruct pulses from the measured data.

## Acknowledgements
Portions of this work were supported by the National Institutes of Health (EB002019).

## Author contributions
P.S. performed simulations and experiments and analyzed the data. W.F. contributed to understanding the pulse evolution. F.W. guided the research. All authors contributed to writing the paper.

## Conflict of interest
The authors have submitted a patent application on gain-managed fiber amplification.

## Supplementary information
is available for this paper in a separate file.

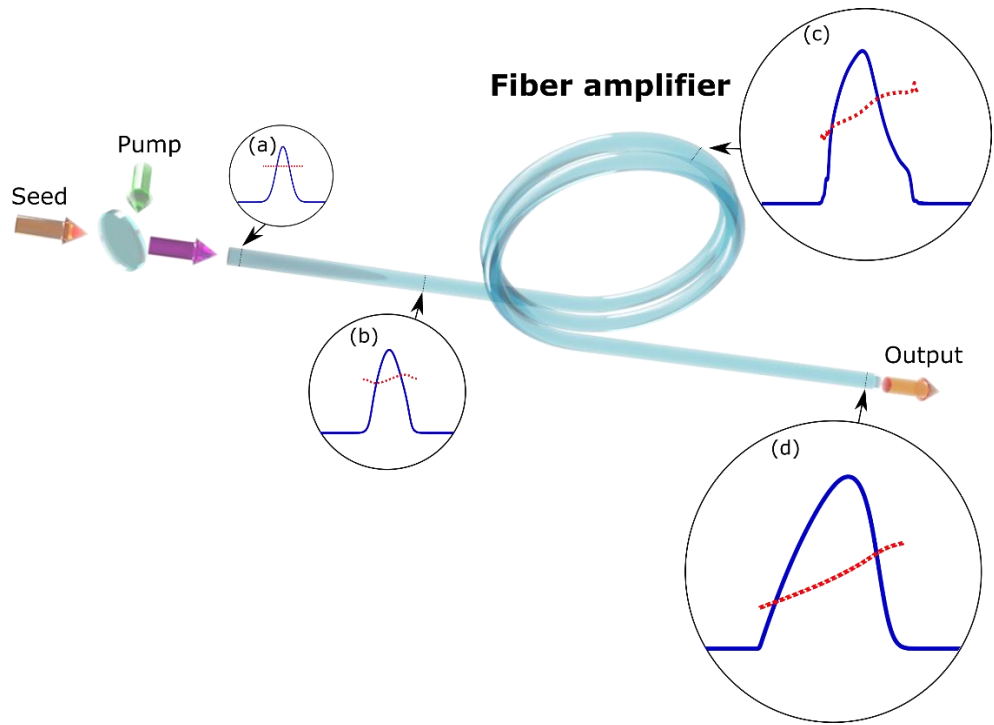

Fig. 1. Simulated example of GMN amplification regime. The temporal intensity (solid blue line) and instantaneous frequency (dotted red line) are plotted for (a) the seed pulse, (b) the self-similar pulse, (c) the pulse beyond the self-similar regime, and (d) the GMN regime.

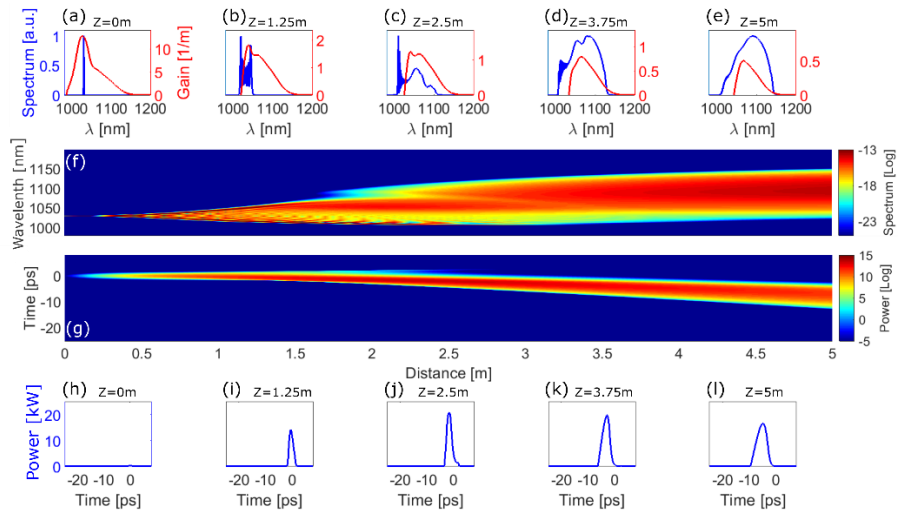

Fig. 2. Simulated pulse evolution in Yb-doped fiber. The top two rows (a-f) show the spectral evolution of the amplified pulse (blue) and saturating gain spectrum (red), while the bottom two rows (g)-l) show the pulse's temporal evolution.

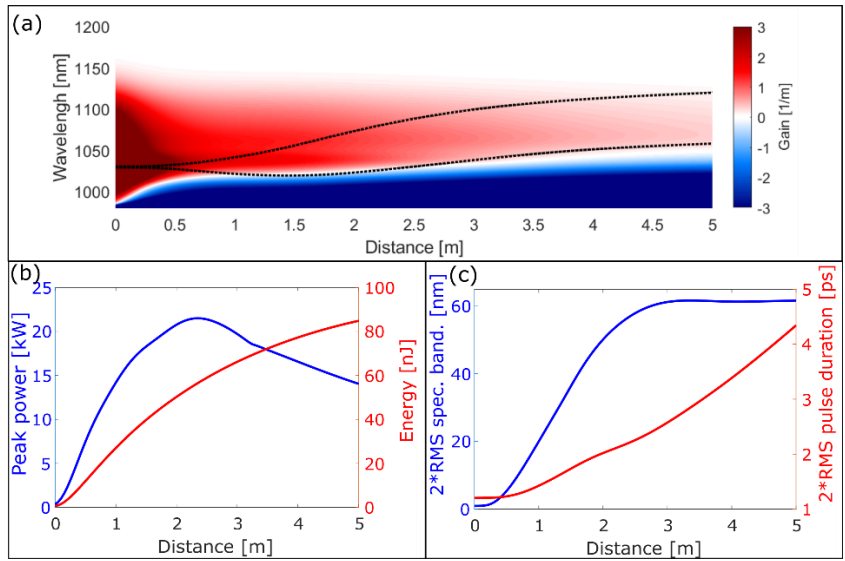

Fig. 3. Pulse and gain evolution in a GMN amplifier. (a) Longitudinal evolution of the gain. Black dashed curves mark the pulse's spectrum (central wavelength ± the root-mean-square bandwidth). (b) Peak power (blue) and pulse energy (red) vs. propagation distance. (c) Bandwidth (blue) and chirped duration (red) of the pulse vs. propagation distance.

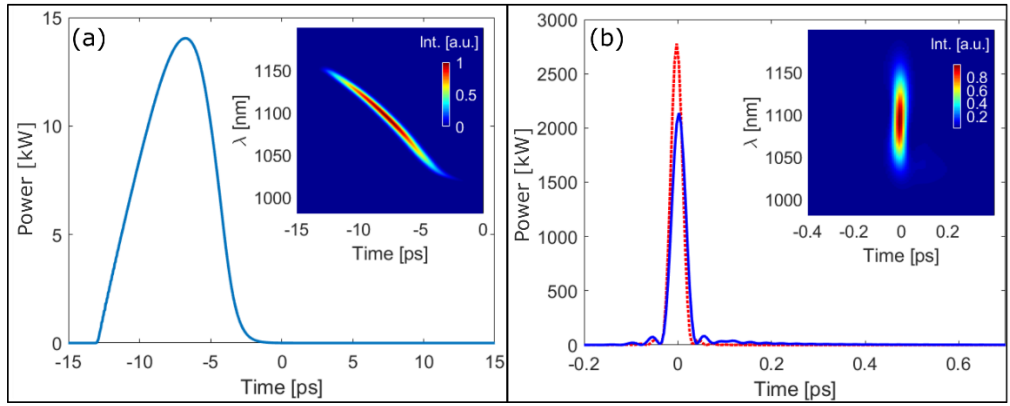

Fig. 4. Pulses generated by a GMN amplifier. (a) Chirped output pulse from the example presented in Fig. 1. (b) Compressed pulse (solid blue curve) and transform-limited pulse (red dashed curve). The insets in (a) and (b) show spectrograms of the chirped pulse (gated by 1-ps Gaussian window) and the compressed pulse (gated by 5-fs Gaussian window), respectively.

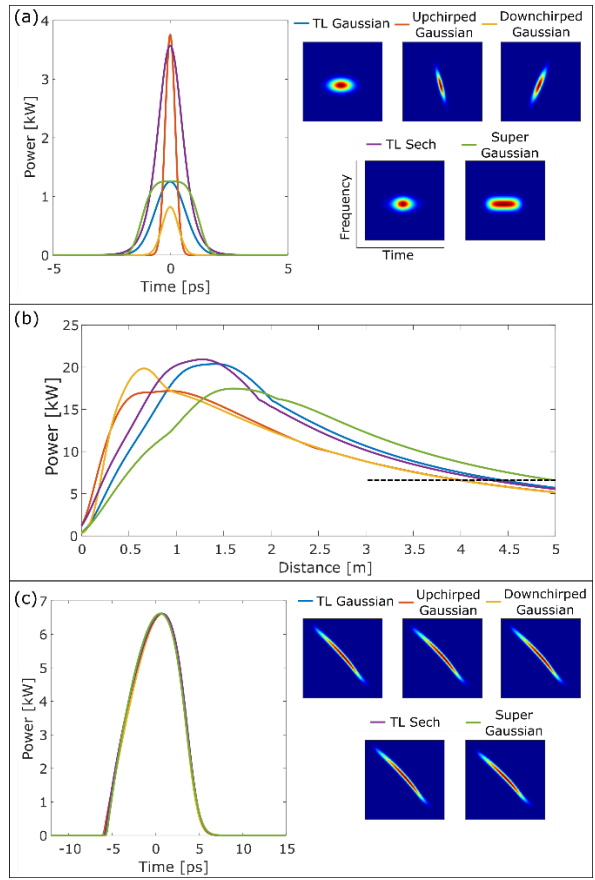

Fig. 5. Numerical evidence of the nonlinear attractor. (a) Different seed pulses and their corresponding spectrograms. TL: transform-limited. (b) Evolution of the peak power for each seed. (c) Amplified pulses produced with the different seeds, taken at the same power level (indicated by the dashed black line in (b), arbitrarily chosen to be 6.6 kW).

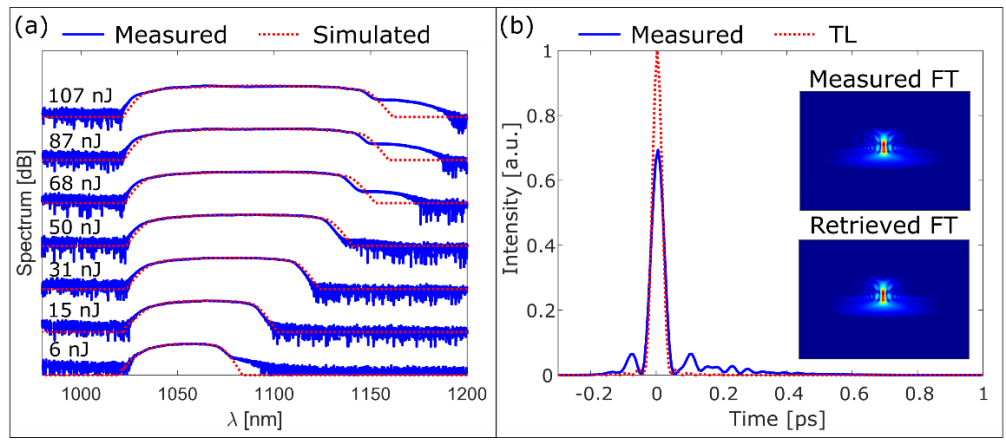

Fig. 6. Experimental demonstration of a GMN amplifier. (a) Measured (blue solid curve) and simulated (red dashed curve) output spectra for increasing pump power (labeled with the output energies). (b) Compressed (blue solid curve) 107-nJ pulse and transform-limited pulse (red dashed curve). Insets in (b) show measured and retrieved SHG-FROG traces.

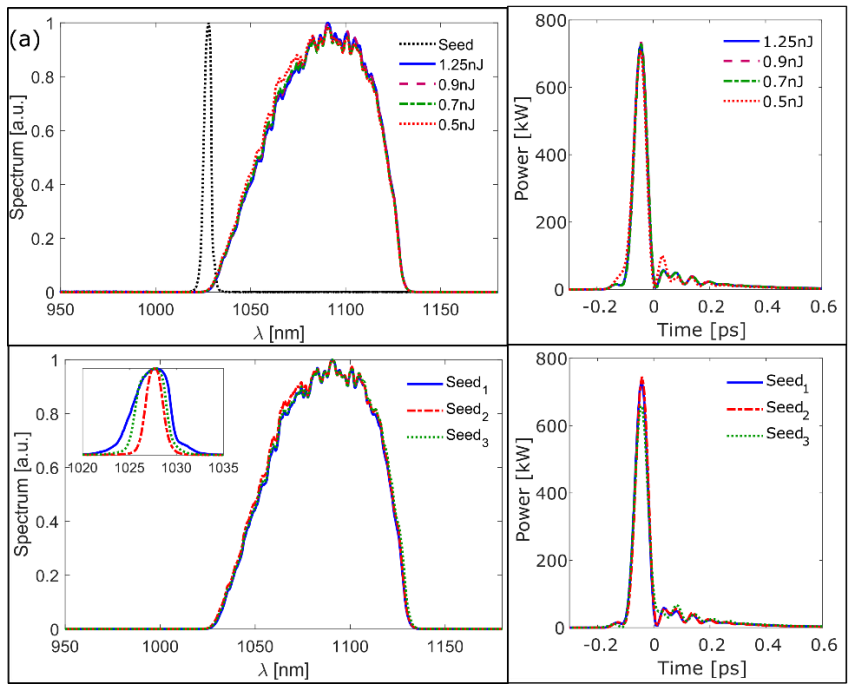

Fig. 7 Experimental evidence of a nonlinear attractor. (a) Output spectra of a GMN amplifier and (b) the corresponding compressed pulses for a constant seed spectrum (black dashed curve in (a)), with the indicated seed pulse energies. (c) Output spectra and pulse shapes of a GMN amplifier seeded with constant seed energies (inset: seed spectra), and (d) corresponding dechirped pulses.

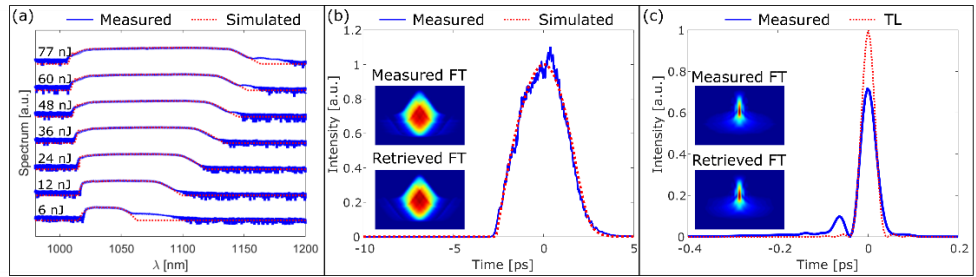

Fig. 8. Direct measurement of chirped pulses from a GMN amplifier. (a) Measured (solid blue curve) and simulated (red dashed curve) spectra for increasing pump powers. (b) Measured (blue solid curve) and simulated (red dashed curve) pulse shapes that correspond to the 77-nJ result in (a). (c) Measured compressed pulse (blue solid curve) and calculated transform-limited pulse (red dashed curve).

# SUPPLEMENTAL INFORMATION
# for
# Nonlinear ultrafast fiber amplifiers beyond the gain-narrowing limit

**Pavel Sidorenko,**[*] **Walter Fu, and Frank Wise**

*School of Applied and Engineering Physics, Cornell University, Ithaca, New York 14853, USA,*
*\*Corresponding author, E-mail: ps865@cornell.edu*

**S1.** Here we analyze the discrepancies between numerical simulation and experimental results presented in Fig. 5. Experimentally, the energy of the amplified pulse is limited by stimulated Raman scattering, visible on the long-wavelength edge of the measured spectra for pulse energies of 68 nJ and above (solid blue curves in Fig. 5). In our numerical simulation, we adde shot noise with one photon per frequency bin to the seed pulse[1]. In this case, our simulations predict that the pulse energy could be increased (by increasing pump power) to 160 nJ [Fig. S1 (a)] before we start to observe significant Stokes waves. To understand this discrepancy between simulations and experiments, we perform several simulations where we vary the amount of shot noise in the seed pulse.

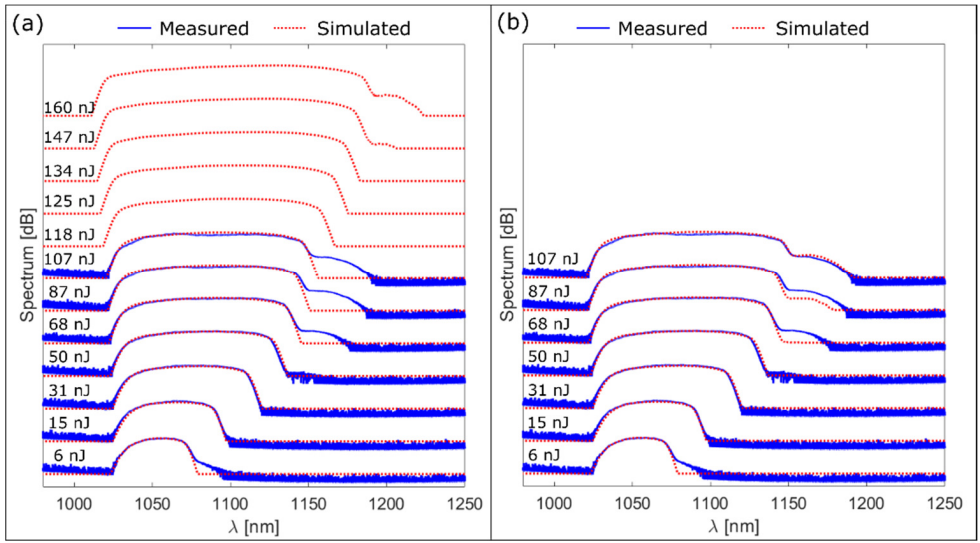

Fig. S1. SRS effects in experimental and simulated GMN amplifier. (a) Compression of experimental results with simulations where 1 photon per frequency band shat-noise was added to the seed pulse. (b) Compression with simulations where 1000 photons per frequency band shot-noise were added to seed pulse. Blue solid curves show experimental results and red dashed curves show results from numerical simulations.

We find that if we increase the shot noise to 1000 photons per frequency bin, the simulated spectra agree with the experimental results [Fig. S1 (b)]. One reason for this may be the generation of amplified spontaneous emission near 1150 nm at the beginning of the gain fiber.

**S2.** Here we experimentally show that pulse energies above 200 nJ are available from a 10-µm-core fiber amplifier based on the GMN regime. In addition, we show that a GMN amplifier can be pumped at different wavelengths. In particular, we show that pumping a GMN amplifier at 915 nm leads to slightly more energetic and more broadband outputs. Figure S2 shows experimentally measured spectra for increasing pump power from a GMN amplifier. For this amplifier, we used 5 meters of Nufern PLMA-YDF-10/125-VIII fiber and 0.7-ps transform-limited seed pulses centered at 1028 nm. The amplifier was co-pumped either with a 976 nm diode (blue curves in Fig. S2), or with a 915 nm diode (red curves). The maximum measured pulse energy was 215 nJ with 915 nm pump and 181 nJ with 976 nm pump. Under both pump wavelengths, the maximal pulse energy was limited by production of Stokes waves.

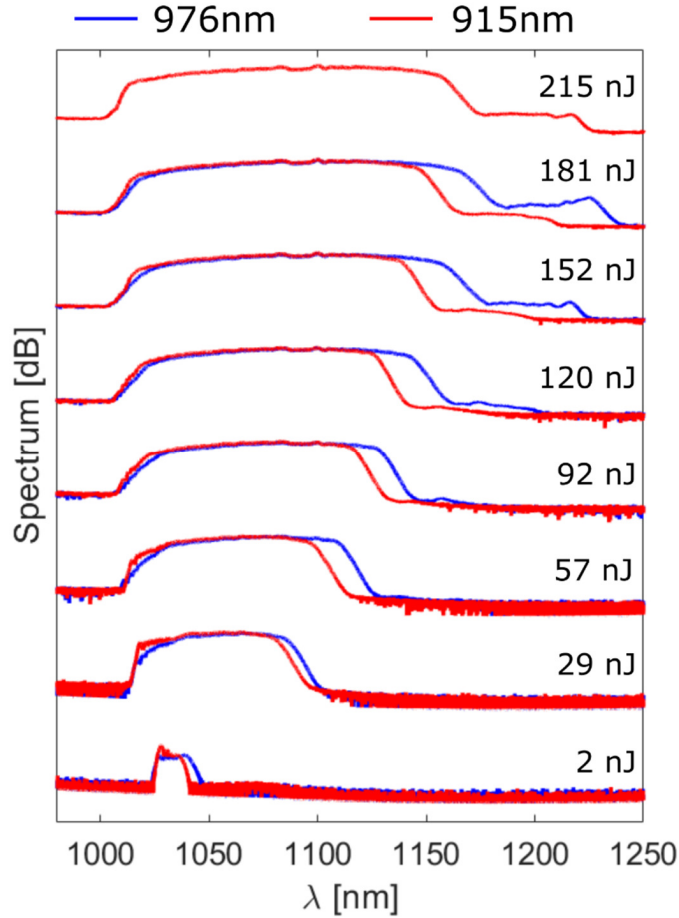

Fig. S2. Experimental demonstration of GMN amplifier. Blue curves show spectra obtained with 976 nm pump, and red curves show spectra obtained with 915 nm pump.

**S3.** Figures S3-S5 show a numerical example of a GMN amplifier using Er-doped gain fiber. In this example, a 1-nJ, 0.8-ps, transform-limited, sech-shaped seed pulse centered at 1520 nm is launched into an Er-doped fiber amplifier that is co-pumped at 980 nm. Here, we simulate a highly-doped Er fiber (RightWave® EDF150 from OFS), where $\beta_2$=36 ps$^2$/km and $\beta_3$=0.07 ps$^3$/km. During the first meter of the gain fiber, the pulse evolution is dominated by SPM [Figs. S3(a-b)]. As in the example with Yb gain, the pulse evolves to a smooth, asymmetric temporal shape instead of experiencing the expected wave-breaking distortions. The longitudinal evolution of the gain spectrum is shown in Figs. S3(a-e) by the solid red curve.

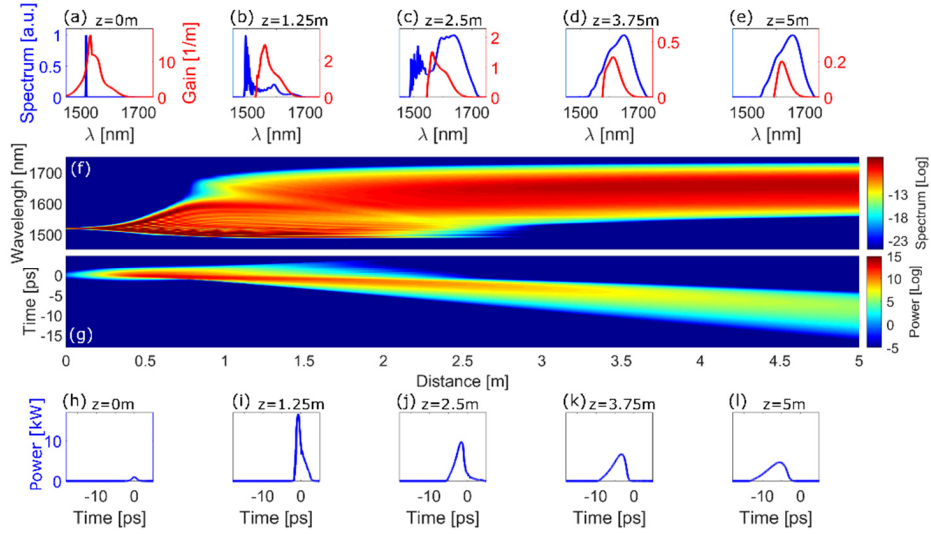

Fig. S3. Simulation of pulse evolution in Er-doped fiber. Top two panels (a)-(f) show spectral evolution of the amplified pulse and saturated gain spectrum. Bottom two panels (g)-(l) show the temporal evolution of the amplified pulse.

Figure S4 (a) shows the evolution of the gain/loss in the amplifier. The black dashed curves mark the root-mean-square bandwidth of the pulse spectrum. Figures S4 (b-c) show the evolution of the pulse parameters.

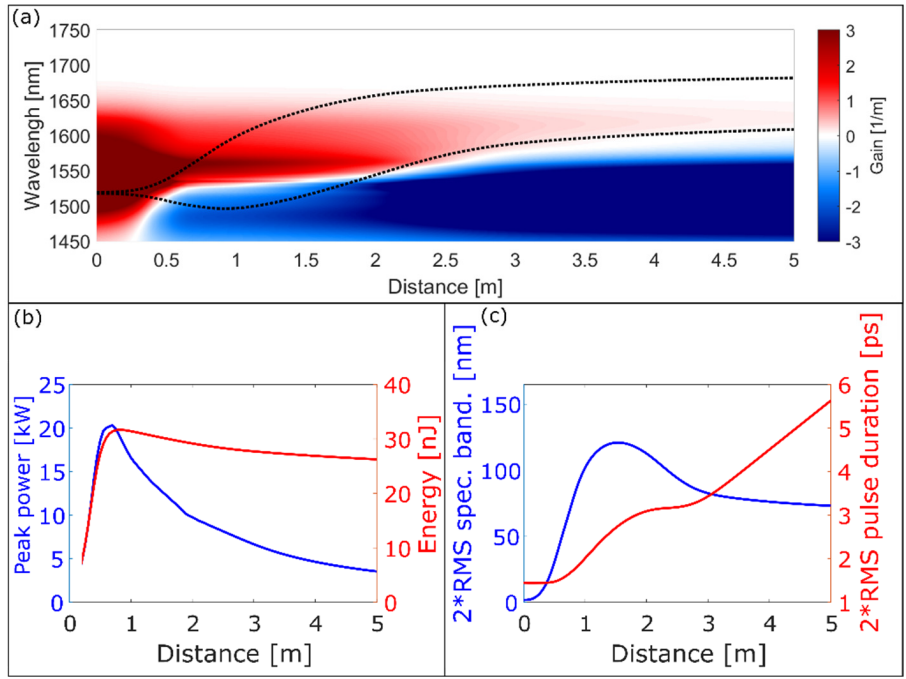

Fig. S4. Pulse and gain evolution in GMN amplifier. (a) Longitudinal evolution of the gain. Black dashed curves mark the spectrum (central wavelength ± the root-mean-square bandwidth). (b) Peak power (blue) and pulse energy (red) vs. propagation distance. (c) Bandwidth (blue) and duration (red) of the pulse vs. propagation distance.

Figure S5 (a) shows an asymmetric, steady-state pulse shape at the output of the amplifier. Figure S5 (b) shows pulse compressed with a grating compressor (solid blue curve). The duration of the compressed pulse is 61 fs.

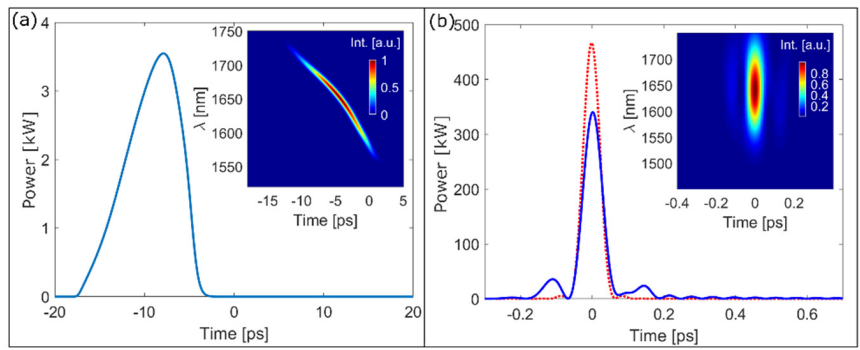

Fig. S5. Pulses generated by an Er-doped GMN amplifier. (a) Chirped output pulse from the example presented in Fig. S3. (b) Compressed pulse (solid blue curve) with full-width-half-max duration 61 fs, and transform-limited pulse (red dashed curve) with full-width-half-max duration 57 fs. Insets in (a) and (b) show spectrograms of the chirped pulse (gated by 1-ps Gaussian window) and the compressed pulse (gated by 5-fs Gaussian window), respectively.

While our results from numerical simulations are promising, we note that the quasi-three level system is less accurate for the Er gain model[2], and numerical predictions remain to be verified by experiment.

**S4.** Here, we show numerically that gain-managed-nonlinearity underlies the pulse evolution in recently-developed, high-performance, Mamyshev fiber oscillators[3,4]. To understand the pulse evolution in the Mamyshev oscillator, we reanalyzed our numerical simulations of a previously-presented cavity[4]. We consider the laser from Ref. 3. Figure S6 (a) shows a schematic of the laser. Figure S6 (b) shows the measured (blue solid curves) and simulated (red dashed curves) spectra for increasing pump power. Figure S6 (c) shows the spectral evolution in the laser cavity of the simulation for 190-nJ output pulses. To understand the pulse evolution in the Mamyshev oscillator, we highlight the spectral evolution in the last gain segment (G2) of the laser cavity [Fig. S7(a)]. As shown in Fig. S7(a), the spectral pulse evolution closely resembles that presented in Fig. 1f. During the first meter of the gain fiber, the pulse evolution is dominated by SPM, and afterwards, the pulse begins evolving toward a smooth spectral shape. Similarly, Figs. S7(b-c) show evolution of the pulse parameters in the gain segment.

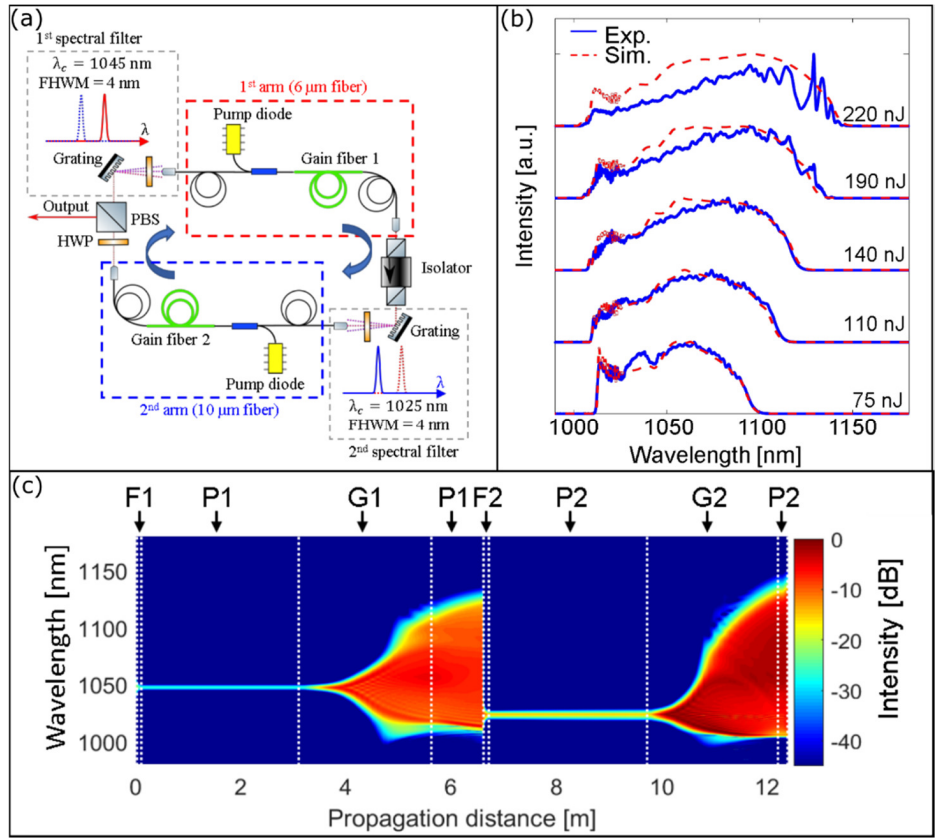

Fig. S6. Pulse evolution in a ring Mamyshev oscillator, reproduced from Ref. 3. (a) Schematic of the laser cavity from Ref. 3. (b) Experimentally measured (blue solid curves) and simulated (red dashed curves) spectra for a range of pulse energies. (c) Spectral evolution of the pulse in the cavity for 190-nJ output pulses.

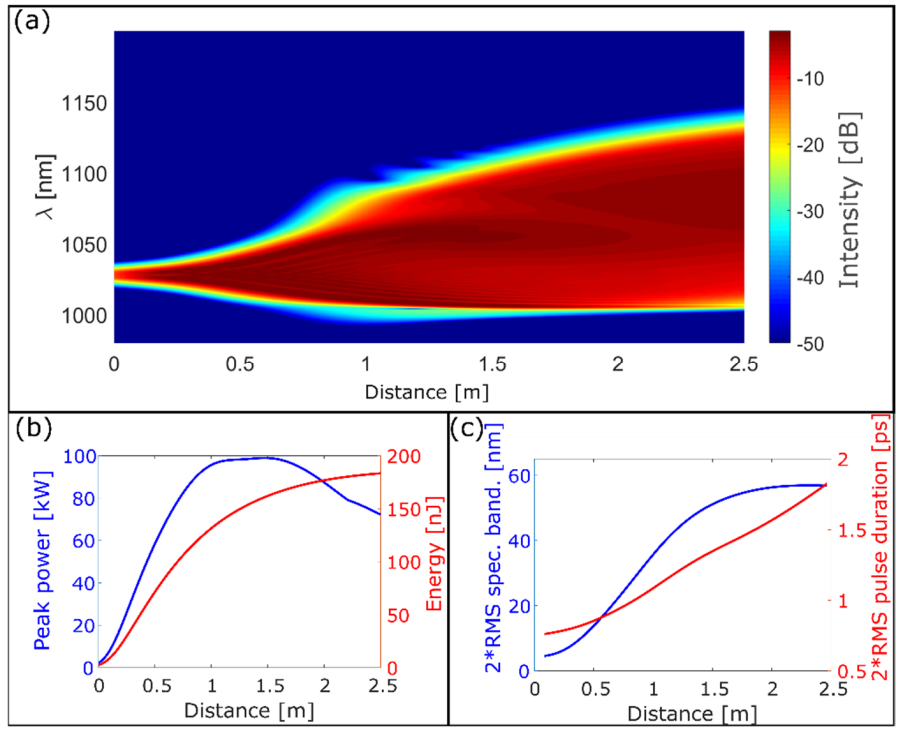

Fig. S7. Pulse evolution in the last gain segment of the ring Mamyshev oscillator. (a) Spectral pulse evolution. (b) Peak power (blue) and pulse energy (red) vs. propagation distance. (c) Bandwidth (blue) and chirped duration (red) vs. propagation distance.

**S5.** In addition to enabling the generation of pulses much shorter than the gain-narrowing limit, GMN amplifiers could be used to realize energetic, ~100-fs pulses in spectral regions not traditionally covered by Yb. Such wavelength-tunable pulses may serve a crucial role in driving hyperspectral nonlinear microscopes[5] or mid-infrared sources[6,7]. Figure S8 shows that the extremely broadband amplified spectrum can be filtered to generate 200-fs, 5-nJ pulses tunable from 1020 nm to 1140 nm. Simulations suggest that the GMN amplifier output spectrum can be extended to allow tuning to nearly 1250 nm, making this amplification regime even more attractive to applications.

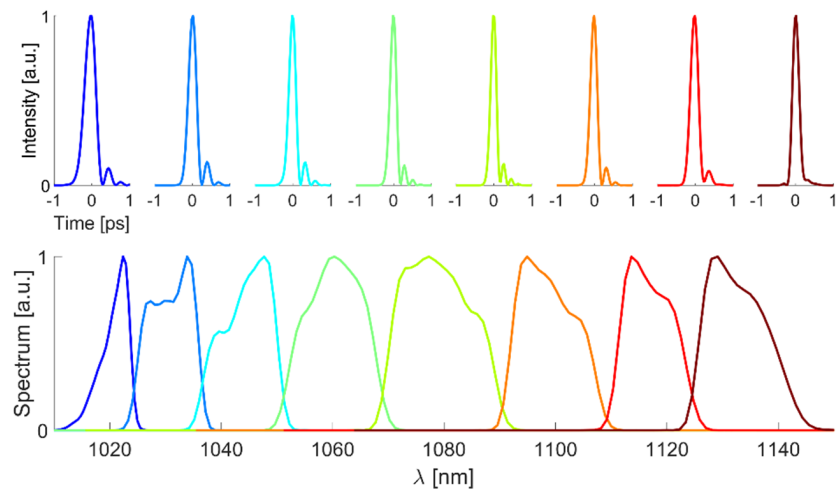

Fig. S8. Experimental demonstration of tunable pulse generation from a GMN amplifier. Top and bottom panels show measured pulses and corresponding spectral slices from the output of a GMN amplifier.